\documentclass{article}
\usepackage{epsfig}
\def\nubar{\overline {\nu} }

\begin{document}     

\title{The east--west  effect for atmospheric neutrinos}
\author{Paolo Lipari  \\
~~ \\
Dipartimento di Fisica, Universit\`a di Roma ``la Sapienza",\\
and I.N.F.N., Sezione di Roma, P. A. Moro 2,\\ I-00185 Roma, Italy \\
~~ \\
also at: Research Center for Cosmic  Neutrinos, \\
 ICRR, University of Tokyo \\
Midori--cho 3--2--1, Tanashi-shi, Tokyo 188-8502, Japan}

\date{March 2, 2000}          
\maketitle

\large

\begin{abstract}
The  atmospheric neutrino  fluxes observed by underground detectors 
 are  not symmetric for rotations  around the  vertical axis.
The  flux is largest (smallest) for the  $\nu$'s  traveling toward east 
(west). This  asymmetry  is  the result of the  effects  of the  geomagnetic
field  on the  primary cosmic rays  and on their  showers.
The size of the  asymmetry  is  to a good approximation 
independent from the existence  of neutrino oscillations, and
can in principle be  predicted  without any ambiguity
due to  existence of unknown new physics.
We show  that there is  an interesting hint of  discrepancy  between the 
experimental  measurement of the  east--west asymmetry by 
Super--Kamiokande and the existing theoretical predictions.
We argue that the  discrepancy is a real  effect caused 
by the  neglect, in the existing calculations,
of the  bending of  the  charged  secondary particles  (in particular
of  the muons)  in the  cosmic  ray  showers. 
The inclusion of this effect in the  calculation
gives  results  in  quantitative   agreeement with the data.
We  comment on the implications of this  result for the study
 of neutrino oscillations.
\end{abstract}

\section {Introduction}
The precise  calculation of the   fluxes of atmospheric neutrinos
has become  a much more  interesting task since
the measurements  of Super--Kamiokande \cite{SK} and other detectors 
\cite{Kamiokande,IMB,Soudan,MACRO} have 
given  convincing  evidence  that    $\nu$ flavor  transitions exist
and are    modifying   the intensity 
and  distorting
the  angular  distributions of  the $\nu$ fluxes.
As the field  of  atmospheric  neutrino is maturing from the 
``discovery  era'' to the era of ``precision measurements''   the  need for 
detailed,  accurate   predictions  for the  the atmospheric neutrino fluxes
has  clearly grown in importance, 
since  comparison  of data  and predictions is  now used 
to  extract  from the data the parameters
that describe exciting new physics. At the same  time
the task of   calculating the  expected  fluxes is  now more 
demanding, 
since  errors in the calculation, or the  use  of flawed data  as  input
can result in the  introduction
of biases in the estimate of these parameters.

The study of the zenith angle distributions of the  neutrino fluxes
is obviously of  central importance in the determination
of the   oscillation parameters.
 The neutrino pathlength $L$
(and  the    density profile  encountered  by a neutrino
during its path)
 are strictly correlated
with the zenith angle $\theta_\nu$
(the correlation is  not perfect  because  the   neutrino  creation
point along  the   straight line   defined
by its momentum at detection is  not exactly known).
The neutrino  flavor transition  probabilities
depend on  the   pathlength  $L$ (or more in  general on the  neutrino path),
and therefore leave their  imprints on  the zenith  angle distributions.

The azimuth angle  distributions are much less interesting.
The only  significant source of  non--flatness  for these distributions
are the effects of  the  geomagnetic field.
In fact, neglecting the existence of the magnetic field,
the geometry of the  neutrino source 
volume (a  sherical shell of air) and  of the  volume  where  neutrinos
propagate,     with  very good
approximation have   cylindrical  symmetry  for   rotations
around the vertical axis\footnote{
Other sources of  asymmetry
can   in principle be the presence of  mountains 
 (especially above the detector)
and the existence of different  profiles  of the atmosphere in
different  geographycal  locations. The  first effect
is easily calculable.  Both  effects  are expected to
be  negligibly small.
}.
Neutrino  oscillations  (or  in fact any other forms
of   transition or  disappearance  proposed as
an explanation for the atmospheric  $\nu$  data)
do not disturb this symmetry\footnote{
If an east--weast asymmetry   is generated by some other source
(for example  geomagnetic effects), neutrino oscillations
can in principle  affect the {\em size} of the  asymmetry, since 
oscillations can distort the $\nu$ energy spectrum. This  effect
is  of  second  order, and a small correction to the calculated
east--west asymmetry.}.
The  fact that the azimuth  angle distributions  are independent
from the  new physics  that  is  investigated, has   
however a  positive consequence, their study
can provide an important   
cross check  for the theoretical predictions  and the experiments.
The calculations must be able to predict  the  shape  of these distributions,
without the need  of additional (unknown) physics  beyond
the standard model,
and  experiments  must be able to measure  
 angular effects  that are unambiguously  predicted.

\subsection {The hint of a discrepancy  between  data  and prediction}

The fluxes   of (positively  charged) primary cosmic  rays  
that  reach the Earth's  atmosphere 
contain  more   particles  traveling  from  west toward  east
than  in the opposite direction.
The discovery of this  effect in  1933  \cite{ew-discovery,Alvarez-Compton} 
allowed to   determine   that  the dominant  component
of the cosmic rays is positively charged.
This  ``primary asymmetry'' is  also reflected in the intensity
of the fluxes of secondary particles  that are 
generated in the showers
of the primary  particles and have  a direction
correlated   with the  primary one.

The Super--Kamiokande (SK) detector \cite{SK-eastwest}
has   measured   the   azimuth angle  distributions of
atmospheric  neutrinos,   finding as expected that the 
distributions  are not  flat, and that there in an excess
of particles  traveling toward east.
In the SK  analysis 
the   east--west  asymmetry is defined as:
\begin {equation}
A = {(N_E- N_W) \over (N_E + N_W) }
\end{equation}
where  $N_E$  ($N_W$)  is  the 
number  of events   with the detected  charged  lepton traveling  
  toward east (west).
After  selecting  events  with the 
charged  lepton ($\ell = e,\mu$) in  the  momentum interval
$p_\ell({\rm GeV}) = [0.4,3]$ and  the  zenith  angle region
$|\cos \theta_\ell| < 0.5$, 
the measured asymmetries are:
\begin{equation}
A_e^{\rm SK} = 0.21 \pm 0.04, ~~~~~~~~ A_\mu^{\rm SK} = 0.08 \pm 0.04
\label{eq:as_SK}
\end{equation}
This has  to be compared with  predictions
from Honda, Kasahara, Kajita and  Midorikawa 
(HKKM) \cite{HKKM} and the Bartol group 
\cite{LSG}:
\begin{equation}
A_e^{\rm HKKM} = 0.13  ~~~~~~~~ A_\mu^{\rm HKKM}  = 0.11
\label{eq:as_honda}
\end{equation}
\begin{equation}
A_e^{\rm Bartol} = 0.17  ~~~~~~~~ A_\mu^{\rm Bartol}  = 0.15
\label{eq:as_bartol}
\end{equation}
Considering  the statistical  errors and the spread in the theoretical
prediction, the SK  collaboration concludes  that  both 
measured asymmetries  are compatible with  the expectations.
However, if we  consider the ratio  $A_e/A_\mu$,
we can  see that there  is  a hint  (at the level
of  2--2.5 standard deviations)  of a  discrepancy between
the calculations   that    predict  asymmetries  that are
approximately  equal  for $e$--like  and $\mu$--like events
 (with only a small excess for $e$--like events)  and  the data   where 
$A_e$ is    events is  
more than  twice  larger that  
the asummetry  for $\mu$--like events.
We also note that there is  an indication that 
the  measured  asymmetry for 
$e$--like events is larger  than the prediction while  on  the contrary
the prediction for $\mu$--like events is  smaller.

A discrepancy between  data and  calculation for the azimuthal 
distributions   can    have a  non negligible importance a
for the interpretation of the data.
The study of the east--west aymmetry 
probes  the  atmospheric neutrino fluxes   in the angular region 
around the horizontal plane
that is actually the most important one for the  determination 
of $\Delta m^2$.
The fact that the  results  for    $e$--like and $\mu$--like events
differ  from the expectations, and in {\em opposite} directions,
in a situation where $\nu$ oscillations 
cannot play a role,  is clearly important for the studies of
neutrino oscillations, where 
the comparison of the events  rates for the two lepton  types
is  a cornerstone of the analysis.

In this  work we predict that with increasing  statistics
the Super--Kamiokande  measurement  of the east--west asymmetry will
clearly show to be in disagreements with the predictions
of  \cite{HKKM} and \cite{LSG},  and indeed  with any 
other prediction 
based on  a calculation   that does  not take  into
account the effects  the  bending in the geomagnetic  field
 of  secondary charged   particles
in the cosmic  ray  showers
 (even  if the calculation  is three  dimensional \cite{fluka-3d}).
The  existing   calculations   predict  an east--west asymmetry
that  is simply the reflection of the asymmetry
of the primary cosmic ray. 
In this case the asymmetries for   $e$--like and $\mu$--like events 
(and indeed for   all for neutrino types
$\nu_e$,
$\nubar_e$,
$\nu_\mu$,
$\nubar_\mu$).
As we will discuss in the following  the effect of the magnetic field
in the  development of the shower  in the atmosphere
results in different asymmetries for the different  neutrino  types,
in good  agreement with the Super--Kamiokande data.

This  work is  organized as  follows.
In the next section  we  rapidly  summarize the  effects
of the geomagnetic  field  on  the  primary cosmic rays.
Section 3   contains a   discussion of the effects of the geomagnetic
field on the  shower development.
Section 4 discusses the  effects of the bending of the  muon
trajectories in 
in the geomagnetic field.
This is the key mechanism that   enhances or suppresses the
asymmetry for the different neutrino flavors.
In section 5  we show  the results  of a complete
calculation  that includes  the  bending  in the  geomagnetic  fields
of all secondary  charged  particles.
Section  6  gives  a summary and some conclusions.
An appendix 
 contains a brief discussion on  the  difficulty of 
performing a   detailed  and  accurate ``three--dimensional''
 calculation of the atmospheric neutrino fluxes
and on the possible  strategies  that are under study.

\section {The East--West effect for  the primary cosmic  rays}
The flux of  primary cosmic  rays  protons  and 
  nuclei  that  arrive  in the vicinity
of the Earth  surface   exhibits the well known east--west  effect,
that is there are more particles  traveling  from  west toward   east
than in the opposite  direction.  This    is  due
to the effects  of the geomagnetic  field  that
forbids the lowest  rigidity particles
from reaching the   Earth.
The effect is  strongest for west--ward going positively  charged  particles.

As an illustration let us approximate the geomagnetic field
as  a  dipole, and study the  trajectories of charged  particles  in  the
equatorial plane. 
The magnetic field  is orthogonal  to the plane,
pointing    ``up''  toward  the magnetic pole in the  northern hemisphere.
Near the surface of the Earth  the field  has  an approximately constant value
$B \sim 0.31$~Gauss. A positively (negatively)  charged  particle  with 
unit charge and  momentum $p \simeq 59$~GeV  can  have a
a clockwise (counterclockwise)   circular trajectory  that 
 grazes  the Earth along
the equator  going from east to west
 (from west to  east)\footnote{The  trajectory
is  unstable  even in the case of an exactly dipolar field}. 
 It  is clear  that an  observer   at the equator  looking    near the horizon
toward  magnetic  east    (west)
can only  see    unit charge  
 primary particles      with a momentum 
larger than   59 GeV,  all  lower momentum particles 
 have ``forbidden  trajectories''.
If the equatorial  observer 
turns  around by   180 degrees, he/she  can see primary
 particles of  much lower  momenta.
This is  qualitatively the origin of the east--west effect.

The problem of calculating the geomagnetic  effects on the primary
cosmic  rays (before  they reach the  atmosphere) 
 can be reduced to  the problem 
of  calculating  ``allowed'' and  ``forbidden trajectories''.
The  definition  of  allowed and forbidden  trajectories is  the   following.
Let  us consider a cosmic ray  of  charge $q$, momentum $\vec{p}$
near the Earth surface  in
position $\vec{x}$, and let us study the past  trajectory
of the particle. 
This  study can have  three  results:
\begin{itemize}
\item[(a)]  the trajectory originates from the Earth's surface;

\item[(b)] the  trajectory  remains  confined  in the 
    volume  $ R_\oplus  < r < \infty$   without 
   ever reaching ``infinity'';

\item[(c)]   the particle   in the past   was at  very large distances 
     from  the Earth. 
\end{itemize}
Trajectories belonging to the classes (a) and (b)  are considered 
as ``forbidden'',  because no primary cosmic  ray  particle
can reach  the Earth   from  a large distance
 traveling along one of these trajectories.
All other  trajectories are allowed.

The algorithm  (used in all calculations of the atmospheric  neutrino fluxes)
to  obtain the primary cosmic ray fluxes 
as a function of position and  angle is    based on two steps:
\begin {enumerate}
\item  It is  assumed  that the primary  cosmic rays  at a distance
of one astronomical  unit from the sun are  isotropic, unless they
are  disturbed by the  presence of  the Earth.
\item  The flux  arriving at  an imaginary   surface 
in the  vicinity of  the Earth (the ``top of the atmosphere'')
is  equal to the isotropic primary flux  after  the  
subtraction of all  ``forbidden trajectories''.
\end{enumerate}
The theoretical  basis for the  ``subtraction agorithm'' 
is  solidly motivated
\cite{Lemaitre-Vallarta}, and is  based on two fundamental 
assumptions, that the  cosmic ray spectrum ``in the absence of the Earth''  is  isotropic, 
and that the field  around the Earth is well described by a static  purely magnetic  field.
The Liouville theorem states   that the density of points  
in phase  space volume is  constant. 
The flux  $\phi(\vec{p},\vec{x})$ is in fact  proportional to the  phase space
density for relativistic particles, and
since the  momentum of  a charged  particle in a
magnetic  field is  constant,
then  the  differential  flux  $\phi(\vec{p}, \vec{x}(t))$
along a  particle trajectory is constant.  It  follows  that
if the  flux is isotropic  at large distances
from the Earth, then the flux in a small cone
around  any  trajectory   is  constant and  independent  from 
the position.  If a particle can  reach the Earth, then the  momentum
spectrum, and the angular  distribution   around it are   not deformed.

For  the explicit calculation of   allowed and forbidden  trajectories,
the most accurate method is the so called
``back--tracking method''   that is a straightforward  direct application
of the definition   outlined  above.
Given a detailed map of the magnetic  field 
around the Earth  (for example  see \cite{igrf}) it is a 
straightforward exercise to  
compute numerically the past trajectories of a cosmic
  ray  from a   point   ``just above''
the atmosphere, and test if it corresponds to an allowed or 
forbidden  trajectory.

It is  well  known that 
the   problem of calculating allowed and  forbidden  trajectories can be 
solved analytically  for  
the  special case of a volume that  is  entirely filled 
with an  exactly  dipolar magnetic  field  (there are no class
(a) trajectories in this model).
In this case  the geomagnetic  effects   result in a sharp
rigidity cutoff''\footnote{In a   numerical calculation for  a 
non exactly dipolar field,
 or  even in the  case of a dipole, if  the trajectories that have a segment
inside the Earth are also considered as forbidden, there is 
 not anymore a sharp cutoff,  but in a  narrow  interval 
 allowed and forbidden bands  of rigidity alternate.}.
 For   a given position $\vec{x}$   
and a given direction $\hat{n}$  the trajectories 
of all positively  (negatively)   particles with   rigidity
$R > R_S^+$   ($R < R_S^-$) are allowed
and the  trajectories of all particles  with  
$R < R_S^+$   ($R > R_S^-$) are forbidden
because they remain  confined  to a finite distance from
the dipole center.
The quantity $R_S^\pm(\vec{x}, \hat{n})$
is the St\"ormer    rigidity  cutoff \cite{Stormer}:
\begin {equation}
R_S^+ (r, \lambda_M, \theta, \varphi) =  \left ( {M \over 2 r^2 }
\right )  ~ \left \{  { \cos^4 \lambda_M \over
[1 + (1 - \cos^3 \lambda_M \sin \theta \sin \varphi)^{1/2}]^2 } \right \}
\label{eq:Stormer}
\end{equation}
where we have made use of the cylindrical symmetry  of the problem,
 $M$  is the magnetic dipole moment of  field,
$r$ is the distance from the dipole, $\lambda_M$ the magnetic
latitude, $\theta$  the zenith angle and
$\varphi$ an  azimuth angle  measured clockwise from magnetic 
north. 
The St\"ormer cutoff has  been  of considerable historical importance. It is 
not sufficiently  accurate for a modern   detailed numerical study. 
but it  remains  
a  good approximation  that  contains all 
qualitative  features of an  ``exact''
numerical  calculation, and remains a very valuable tool to gain
 physical insight and qualitative understanding.
We can use it  to ilustrate the three most important 
qualitative  points that we  will need for the discussion  in this  work.
\begin{enumerate}
\item  The  geomagnetic  effects depend  on the detector
position. They are  strongest  near the equator
and  weakest at the  magnetic poles.
In equation (\ref{eq:Stormer})   we can see 
that the value of the cutoff grows monotonically   from  a
 vanishing  value  at the magnetic pole  ($\lambda_M = \pm 90^\circ$)
  to  a  maximum   value at the magnetic  equator  ($\lambda_M =0^\circ$).

\item  The  set  of  allowed  rigidities is largest 
for  particles  traveling  toward magnetic  east.
In  equation (\ref{eq:Stormer}) 
for  a  fixed detector  position (fixed $\lambda_M$)
and  a  fixed zenith angle $\theta$,  the cutoff  is  minimum
(maximum) for  azimuth angle  $\varphi =  270^\circ$ 
($\varphi =  90^\circ$) that corresponds to a particle  traveling toward
east (west).

\item  For a fixed  detector position and 
a fixed  azimuth angle 
the   set of allowed rigidities is largest
(smallest)  for  vertical (horizontal)  particles.
In  equation (\ref{eq:Stormer}) 
for  $\lambda_M$ and $\varphi$ fixed, the cutoff grows
monotonically   with $\theta$.
\end{enumerate}

The east--west aymmetry of the primary cosmic  ray  radiation 
is of course reflected in  asymmetries for the fluxes
of  secondary particles  generated in their showers  in the atmosphere.
This  effect   results 
in asymmetries of  approximately the same size for  all 
four  neutrino types. This   is  a consequence of the fact 
that a  primary  particle of  energy $E_0$     produces
yields  of  $\nu_e$,
$\overline{\nu}_e$, 
 $\nu_\mu$ and
$\overline{\nu}_\mu$ 
with approximately the same   energy spectrum (and   relative normalizations
 in the  ratios  $1 : 1 : 2 : 2$). 
The   existing calculations of  the neutrino fluxes 
only include  the  geomagnetic  effects  on the  the primary  flux
as  a  source of an  azimuthal  asymmetry
and  as  a consequence the    predicted  east--west neutrino asymmetries
are all of approximately the same size.

\section {Effects of the geomagnetic  field  on the shower development}

\subsection {Bending of the trajectories of charged  particles}
The trajectories of all charged particles in  cosmic rays  showers
are curved  because of the  presence of the geomagnetic field.
It will be particulary important in this  work
to consider the effect  of the bending
on the zenith angle of the particles, that is   to study if the
particle  trajectories   are bent  ``upward'' 
  or ``downward''.
We can define a  (position dependent) 
 system of  coordinates  with the $z$ axis
pointing up,  the $x$ axis pointing  toward  magnetic north
and the $y$ axis  completing  a right--handed  system pointing toward the 
magnetic west.
In this  system the  magnetic  field  has  by  definition
components $\vec{B} \equiv (|B_{\rm hor}|, 0, B_{\rm vert})$  
with  $B_{\rm vert}$  positive  (negative) in the
 southern (northern) hemisphere.
The equation of  motion of a charged  particle in the geomagnetic  field
for the   vertical   component is:
\begin{equation}
{dp_z \over dt}  = -q \, \beta_y \, |B_{\rm hor}| = 
 +q \, \beta_{\rm east} \, |B_{\rm hor}|,
\label{eq:bending}
\end{equation}
where $q$ is the  electric charge and $\vec{\beta}$ 
the velocity of the particle.
Positively  charged  particles  ($q >0$)  traveling east
($\beta_y < 0$) have  $dp_z/dt > 0$, and are   bent  up.
Negatively  charged particles
with the same initial  direction are  are  bent down.
The reverse happens for  particle    traveling in the
 opposite  (west--ward) direction  ($\beta_y >0$): negatively charged 
 particle are  bent up  while positively charged ones   are bent  down.

\begin{figure} [t]
\centerline{\psfig{figure=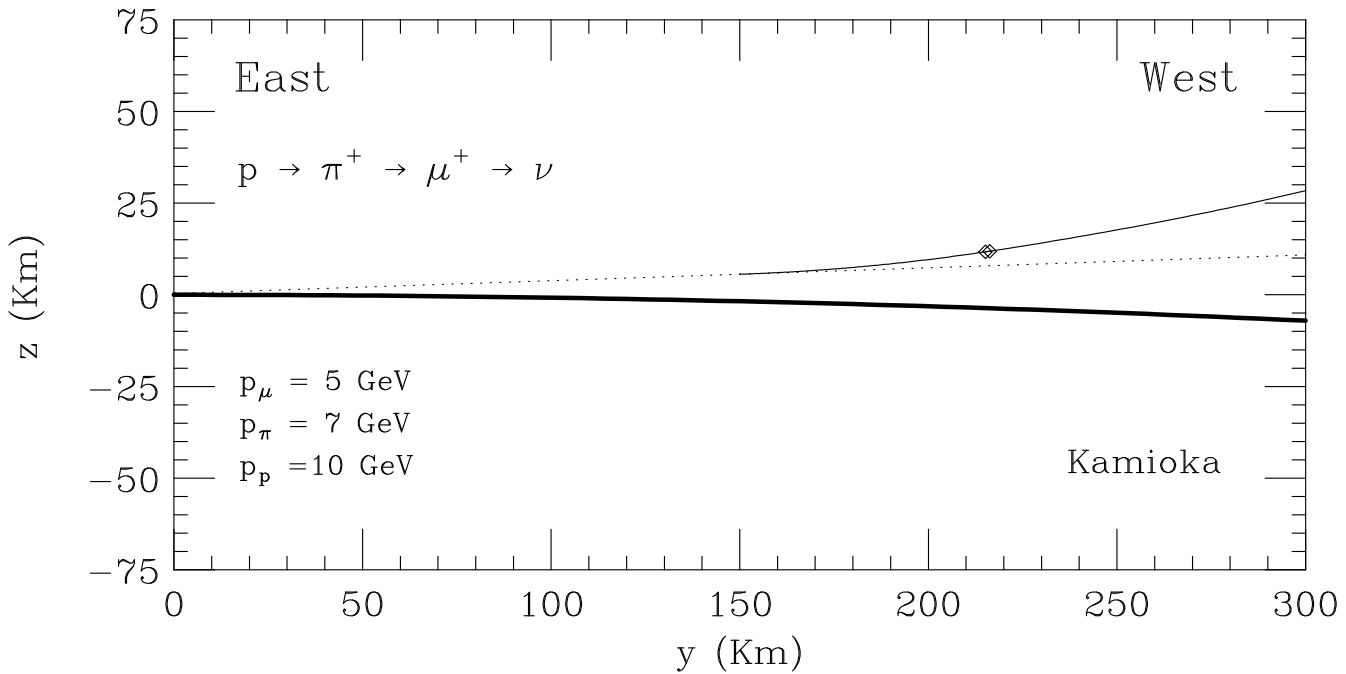,angle=90,height=9.0cm}}
\caption {Deviation of a muon in the geomagnetic  field
above the SK detector (see text).
\label{fig:ew1}  }
\end{figure}

An example of the bending of charged  particles is  shown in
fig.~\ref{fig:ew1}.
The figure is in true  scale  and represents  a projection of  
the  space above the Super--Kamiokande detector.
The center of the SK detector
(latitude 36.42$^\circ$~N,  longitude 137.31$^\circ$~E,
and altitude 371.8 meters above sea level \cite{SK-eastwest})
corresponds in the figuere to   the  point
with  coordinates (0,0.3718).
The thick line  represents the  ``sea level'' surface
of the Earth in the vicinity of SK detector.
The  $z$ axis  corresponds to the vertical
axis  that passes through 
 the SK center, and the horizontal  axis  points to geographical
west.
The thin solid lines describe the projections  in the plane
of the figure of the  trajectories of three charged  particles.
The particles   are part  of  a cosmic ray shower and are 
``connected''  by  a production chain 
$p \to \pi^+ \to \mu^+$.  The trajectories are  calculated  integrating
the equations of motion for charged  particles in 
the International Geomagnetic
Reference Field (IGRF \cite{igrf}) for the year 2000
and correspond to a 10~GeV  proton, a 7 GeV $\pi^+$ and a 5~GeV
$\mu^+$. The pion production and decay points are marked by two diamonds.
The muon  emits  a neutrino (the  trajectory and its continuation
beyond  is indicated by  a dotted line)  that intersects the SK  detector.
For  illustration both the pion  and the muon decay  after  exactly
three proper  lifetimes. Both the pion and the muon are emitted 
exactly collinearly with respect to the primary  (or parent) particle, so 
the angle  between the  neutrino and the primary  directions
is  entirely due to the effects of the geomagnetic  field.
Since  all three 
 particle  are positive  and travel toward the east direction
they are bent upward.

In the following we will study the effects  of the bending
of  the shower particles in the magnetic  field  on the fluxes
of   atmospheric  neutrinos.
These effects have been  ignored
in all  previous calculations.
In our  numerical  work we have  included the    bending of all 
charged particles  in the shower.  In  the  qualitative discussion that 
follows  we  will discuss separately the effect of the binding on the 
primary particles,  and on the muons  produced in the showers.
These are in fact the most important sources    of the effects  we 
have found.
Charged pions  have a  a  lifetime that is two  orders of magnitude
shorter   than $\tau_{\mu}$ (and a mass similar
to $m_\mu$), therefore their  bending in the magnetic  field
is  much less  important   than  for muons.
Most of the neutrinos  are produced in the chain decay of mesons
produced  in the first interaction of a primary.
Also for those produced deeper in  the  shower  most of the 
interesting  effects is   the result of the bending of 
the primary particles and the muons.

Note  that  we can expect that  the effects  of the field will be
especially important for
 the neutrinos  that  come from muon decay, and significantly smaller
for the  muon (anti)--neutrinos  produced  directly in 
meson decay.

\subsection {The  bending  of the primary particles}

It is  incorrect to think that 
all effects   of the magnetic   field on  the primary particles
are    taken into account 
by the  study of allowed  and forbidden  trajectories
discussed in  section~2.
The bending of  the primary particles
is  a factor in  determining   the  average  altitude 
of the primary interactions,
and   the  zenith  angle  distribution 
of the  primaries at the interaction point.
These  two  effects are not taken into  account  in the existing
calculations.
Note that  the zenith   angle 
(that is the polar  angle with respect the local  vertical)
of a particle 
can be calculated  as:
\begin{equation}
\cos \theta = - { \vec{p}\cdot  \vec{r} \over |\vec{p}|~ |\vec{r}| }
\end{equation}
and  is  not a constant even for  particles  traveling
 along a straight line.
The  value of the zenith angle along a particle trajectory
will  of course  be determined by the bending in the magnetic  field.

As an illustration let us
  study  the  trajectories of  two protons  both on  allowed  
trajectories  that cross an imaginary spherical  surface above the Earth
with the same zenith angle but    different  azimuth angles, 
with one   particle   traveling toward  east, and  second one 
  toward west.
If the two  trajectories are  approximated as  straight lines
the    distributions of  altitude  of  the  interaction   point
for the two particles 
(determined by the interaction length $\lambda_{\rm int}$, the 
inclination of the  trajectory and the altitude profile of the air density)
will   of course be identical. 
It is  however easy to see  that   since the  east--going particle is
bent ``up'', and the  west--going one is  bent ``down'',
the first particle will on  average interact interact  higher, and 
more ``horizontally'':
\begin{equation}
 \langle h_{\rm int} \rangle_E >   \langle h_{\rm int} \rangle_W,
\label {eq:hprim}
\end{equation}
\begin{equation}
 \langle \cos \theta_0^{\rm int} \rangle_E  <
  \langle  \cos \theta_0^{\rm int} \rangle_W
\label {eq:ctprim}
\end{equation}
Both the  inequalities  (\ref{eq:hprim}) and (\ref{eq:ctprim})
result in an enhancement of
the east--west  effect for the neutrinos.
The   average number of neutrinos  produced in  a   shower
is larger   when   the   primary  particle interacts  higher 
because the muons  have a longer pathlength 
for decay, and the 
shower  develops  in a medium of  lower  density
where   the decays of  mesons  is enhanced, 
 and muons lose less  energy before  decay.
For  the same  reasons 
more inclined showers  produce  more  neutrinos.

In summary we can expect that the   bending of the primary particle
inside the atmosphere   result in an  enhancement 
of the neutrino east--west asymmetry
with respect to the estimates  that  only consider the bending in vacuum.
This  enhacement
will be approximately equal  for  all  four neutrino
types, it is therefore an interesting effect, but it is cannot be the
explanation  for the 
(hint of) discrepancy  between  data and prediction   represented
by equations
(\ref{eq:as_SK}),
(\ref{eq:as_honda}) and
(\ref{eq:as_bartol}).

\subsection {The angle between the neutrino and the primary particle}
For the effects that we want to consider it  is essential to consider
the difference in  direction between  a neutrino
and the primary  particle  that  generated the shower where   it was 
produced.
We  introduce the notation:
 $\Omega_\nu = \Omega_{0} \oplus \Omega_{0\nu}$
where
 $\Omega_{\nu}$
 is the  direction of the detected  neutrino,
 $\Omega_{0}$ the direction of the primary particle,
and  $\Omega_{0\nu}$  the   angle between  the neutrino  and the primary.
The angle $\Omega_{0\nu}$  is the  result of the combination
of  several processes.
For the neutrinos  that are produced  directly in a meson decay 
we   can write:
\begin{equation}
\Omega_{0\nu} =   \Omega_{0\pi}  \oplus  \Omega_{\pi\nu}
\end{equation}
where 
 $\Omega_{0\pi}$  the direction of emission of the parent pion
in the primary interaction,
and  $\Omega_{\pi\nu}$ the   direction of emission of the neutrino
in the pion decay (this qualitative discussion  obviously applies also
for kaons).
Similarly for  neutrinos  emitted in the chain decay 
$\pi \to \mu \nu$ we have:
\begin{equation}
\Omega_{0\nu} =  \Omega_{0\pi}  \oplus
 \Omega_{\pi\mu} \oplus  \Omega_{\mu B}  \oplus  \Omega_{\mu\nu}
\end{equation}
where
 $\Omega_{\pi\mu}$
  ($\Omega_{\mu\nu}$)  is  the   emission   direction 
of the muon (neutrino) in the decay of the pion (muon),
and $\Omega_{\mu B}$ is  the deviation   of the muon 
in the geomagnetic  field (the deviations  in the  magnetic  field
of the other charged particles are 
  typically two orders of magnitude  smaller;
they are included
in our numerical  work).

The average  deviation (considered as  a space angle) 
$\alpha_j$ for  the process $j$ 
can be easily estimated  in  first order:
\begin {equation}
\alpha_{0\pi} \sim {\langle p_\perp \rangle_\pi \over p_\pi} \sim
 {4 \langle p_\perp \rangle_\pi \over p_\nu } \sim 
 {5.2^\circ \over p_\nu({\rm GeV})},
\end{equation}
where   we have used the   approximation
$p_\pi \sim 4p_\nu$ (since a charged  pion  energy  
is  approximately shared
equally between  three neutrinos  and  a $e^\pm$)
and  have assumed  a  transverse momentum
$\langle  p_\perp \rangle  \sim  350$~MeV;
\begin {equation}
\alpha_{\pi \nu} \sim {p^*  \over p_\nu} \sim
 {1.7^\circ \over p_\nu({\rm GeV})}
\label{eq:theta_pinu}
\end{equation}
where $p^*$ is the  center  of
 mass momentum  of the  final state  particles
in  the decay $\pi \to \mu\nu$;
\begin{equation}
\alpha_{\pi \mu} \sim {p^*  \over p_\mu} \sim {3 \,p^*  \over p_\nu}
\sim  {0.6^\circ \over p_\nu({\rm GeV})}
\end{equation}
this  is  approximately  3 times  smaller that
(\ref{eq:theta_pinu})  because the muons    have on average  three
times the   energy of their neutrino  decay products; 
\begin{equation}
\alpha_{\mu \nu} \sim {m_\mu/3  \over p_\nu} \sim 
 {2.0^\circ \over p_\nu({\rm GeV})}
\end{equation}
and  for the deviation of the muons  in the magnetic field:
\begin{equation}
\alpha_{\mu B} \sim {L_\mu  \over  R_\mu} 
\simeq \left ( \tau_\mu \,{p_\mu \over   m_\mu}   \right )
~ \left ( {e \, B \over p_\mu} \right ) \sim 
10.7^\circ  ~B({\rm Gauss}) 
\end{equation}
where $L_\mu$ is the muon  decay  length,
$R_\mu$ the muon  gyro--radius in the  geomagnetic  field, and
$B$ the value of the field.
In this  estimate  we  have neglected    the muon  energy loss
and  we have  also  assumed that 
the muons   decay before hitting  the ground.

There   are two  fundamental  differences  between  the
effect of  the magnetic bending
and the  other   contributions  to  $\Omega_{0\nu}$.
The first difference  is that all  deviations,  except the
magnetic  bending one  scale 
  $\propto p_\nu^{-1}$  reflecting the
forward boost 
in the interaction or decay of relativistic  particles.
The  deviation of the  muons in the  geomagnetic  field
is (in first order) independent from the muon  momentum,
because  both  the    magnetic   rigidity   and the decay
pathlength   increase   proportionally to $p_\mu$.
Higher  momentum muon   bend  less in the  field, but
live  longer  and the field  can act for a longer time.

The second  and  most  important  difference is that 
all  deviations  (except the bending one)  are azimuthally
symmetric, therefore  if we consider the deviation in any plane,
the    average value vanishes.
The   deviation in  the magnetic  field is of course not 
an ``average    value'', all muon   of the same  initial momentum
have  exactly the same  deviation, and the  deviation is not  azimuthally
symmmetric  but happens  in a well defined   plane.
If we for example consider the average zenith angle of  the neutrinos
produced  in the showers  generated  by  primaries of fixed
direction $\Omega_0 \equiv (\cos \theta_0, \varphi)$  we find
that the average contribution  of  all  sources vanishes, except
for the  effect of the  bending  in  the field.
\begin{equation}
\langle  \theta_\nu \rangle  = \theta_0 +
\langle  \theta_{\mu B} \rangle 
\label{eq:zenith}
\end{equation}
This  effect  plays  a  crucial role  in  determining the
neutrino east--west asymmetry.

\section {Magnetic bending of  $\mu^\pm$ 
  and the $\nu$ east--west asymmetries}
\label {sec:mu}

\begin{figure} [t]
\centerline{\psfig{figure=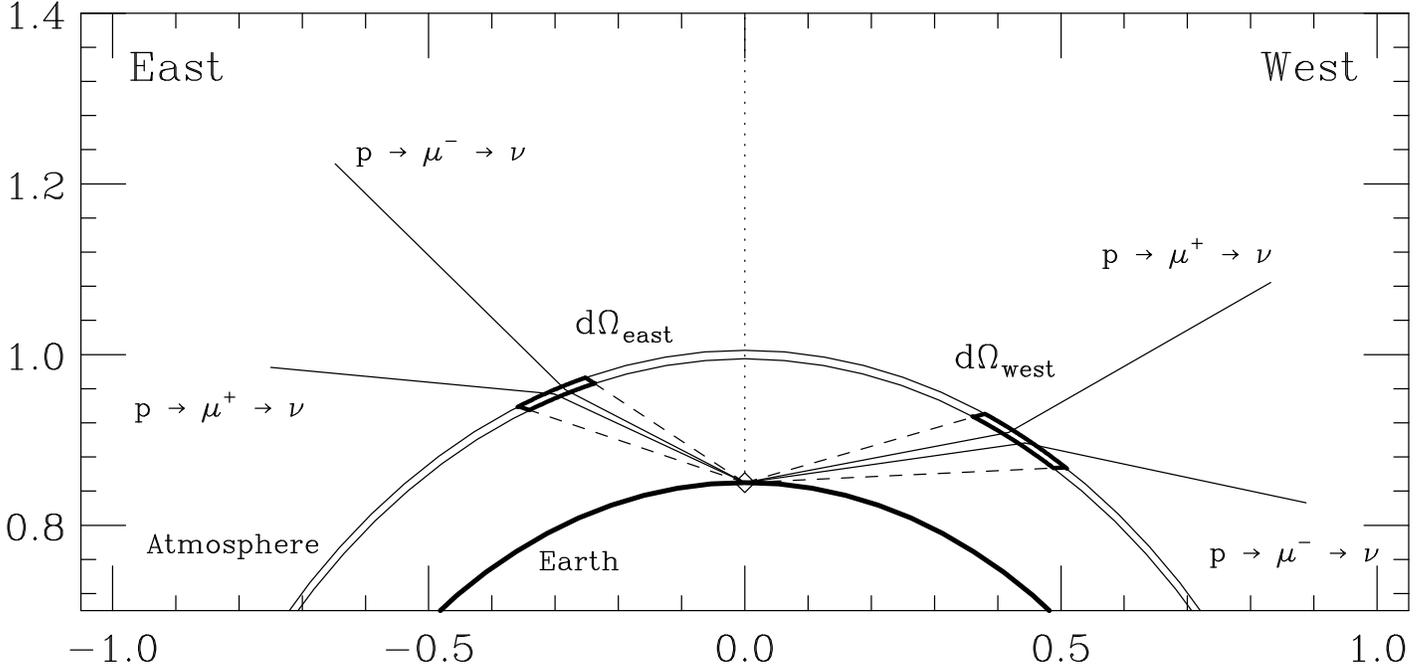,angle=90,height=9.0cm}}
\caption {Relation between  the  directions of  of the primary
particle  and of the  detected  neutrinos,  for  $\nu$'s
produced in  muon  decay  (see text).
\label{fig:ew2}  }
\end{figure}

The effect of the geomagnetic  field on the
trajectory of charged particles   (equation 
(\ref{eq:bending}))  implies
(a  scheme  of  the  results is  also shown in fig.~\ref{fig:ew2}):
\begin {itemize}
\item [(a)]  Positively charged  particle ($q>0$)  traveling 
toward east  ($\beta_y <0$)  are  bent  ``up''.

\item [(b)]  Positively charged  particle ($q>0$)  traveling 
toward  west   ($\beta_y >0$)  are  bent  ``down''.

\item [(c)] Negatively charged  particle ($q>0$)  traveling 
toward east   ($\beta_y <0$)  are  bent  ``down''.

\item [(d)] Negatively  charged  particle ($q>0$)  traveling 
toward  west   ($\beta_y >0$)  are  bent  ``up''.
\end{itemize}

From equation (\ref{eq:zenith})
we can then  deduce that 
neutrinos produced in the decay of
positive (negative) muons  in  showers of 
 east--ward  (west--ward)  traveling primaries
will be  on average more    ``horizontal'' than the  primary  particle;
conversely  the 
neutrinos produced in the decay of
positive  (negative) muons  in  showers of west--ward 
(east--ward) traveling primaries
will be  on average more    ``vertical'' than the  primary  particle.

The key remark is  now, that this systematic difference 
between the zenith angle  of the neutrino and the primary
results for   the cases  (a) and (d)  in 
an  enhancement
and for the cases (b) and (c) 
in a suppression of  the neutrino  flux.

Two  effects  play  a role  in  the  enhancement or  suppression
of the flux.
The  first effect is  purely geometrical and  would  be present
even for  an exactly isotropic primary flux
(this  is  not  so academic  as  it may  sound, because the primary
flux {\em is} actually   isotropic for all   rigidities above 60 GV).
In  this case the rate of 
primary interactions  with  zenith angle  $\theta_0$ is 
$\propto \cos \theta_0$, and  there is an  excess of vertical showers
over horizontal ones.
Therefore   if  the  zenith angle of the primary  particles
is  smaller  (larger)  than the neutrino one  we can  expect
a  larger (smaller) event rate.
This is  illustated schematically in  the  diagram of fig.~\ref{fig:ew2}.
The neutrinos  detected by 
an observer (indicated  by a diamond) located  on the surface  of the Earth 
(shown as a thick line)    looking 
in the solid angle $d\Omega_{\rm west}$  (indicated  by  a dashed cone)
are  produced 
 in  a  ``patch'' (outlined with a thick line)
 of atmosphere 
(that in the diagram is  represented  by a thin spherical shell)
subtended  by  the solid  angle  $d\Omega$.
It can  be seen from   the figure that
this  patch of atmosphere ``looks''  larger 
(it has a larger  projected  area)
for  the   primary  particles  if the   production chain
is $p \to \pi^+ \to \mu^+ \to \nu$   (the neutrinos
can  be $\nu_e$ and $\nubar_\mu$)  
than in the case  when the neutrinos
are  produced  directly   in meson decay.
This  is  a simple  consequence of the  fact that the projected area
of the element of  atmosphere  subtended by the 
solid  angle $d\Omega$    is  proportional to $\cos \theta_0$.
For a fixed  neutrino zenith angle, the  average  
value of  the primary  zenith angle
is  not a constant  but depends  on  the
production mechanism.
In the case 
 $p \to \pi^+ \to \mu^+ \to \nu$  
the relation  between the  zenith angle of the  primary and of
the neutrinos
is  $\langle \theta_\nu \rangle  <  \theta_0$,
in the case of neutrinos   directly from 
meson decay:  $\langle \theta_\nu \rangle  \simeq  \theta_0$.
Similarly the patch of atmosphere  ``looks''  smaller  for  the primaries
when the production process is 
$p \to \pi^- \to \mu^- \to \nu$   (the neutrinos
are $\nubar_e$ and $\nu_\mu$)  since
in this case the  relation 
between the zenith angle of the neutrinos and the 
primary is:  $\langle \theta_\nu \rangle  >  \theta_0$.
The bending in different   directions of  positive  and 
negative  muons  results in one  case  in an  enhancement, and in the 
other in
a suppression of the neutrino  fluxes.
This  discussion  can be  repeated 
for a  solid angle  $d\Omega_{\rm east}$  in the opposite hemisphere,
with the crucial  difference that the  production chains  that are
enhanced  and suppressed are interchanged.

If now  we consider the  fact that  the primary  cosmic  rays  that 
reach the Earth's  atmosphere  are not  isotropic  we
find actually that  the  enhancement and suppression are
{\em stronger}  that  what is obtained  with the geometrical
argument  outlined  above.
This  can  be uderstood  noting  that the set of
set of  forbidden  trajectories    becomes  larger
when the  zenith angle increases.  This can be easily seen from
the St\"ormer formula (\ref{eq:Stormer}) where
for  fixed $\lambda_M$ and $\varphi$ the   rigidity  cutoff
grows  monotonically  with the zenith angle  $\theta$.

In summary:  the  effect of  the   magnetic  bending 
of the muons  results in  an enhancement 
for the flux of  the neutrinos  produced in the decay
of $\mu^+$    (that is  $\nu_e$ and $\nubar_\mu$)
when the particles   travel toward east, and 
a suppression  when the particles travel toward  west.
The net effect is  a {\em positive} contribution to the east--west
asymmmetry.
The effect of magnetic  bending is  obviously opposite 
for the  flux of neutrinos  produced in the decay
of $\mu^-$    ($\nubar_e$ and $\nu_\mu$).
The flux is suppressed when the 
when the particles   travel toward east, and 
enhanced  when the particles travel toward  west.
The net effect is  a {\em negative} contribution to the east--west
asymmmetry.

The effect of the   magnetic  bending of the muons has  to be combined
with the  asymmetry  generated by the geomagnetic effects on the
 primary particles, that to  a  good approximation result in an equal
contribution to the asymmetry of  all  neutrino types.
Moreover,  for  the neutrinos  produced directly in meson  decay, the 
effects of the geomagnetic  field  during the development the shower
are  negligible, and for them  the only  significant 
source of asymmetry are  the geomagnetic  effects  on the primary
trajectory.

\subsection {Qualitative  predictions}
We can  finally  put all   results together 
and  give  some  simple,  non--trivial  and  testable
predictions.

\begin{enumerate}
 \item   With increasing statistics, the Super--Kamiokande detector
will    detect  east--west asymmetries that are {\em not} in agreement
with predictions  based on one--dimensional  calculations.

\item The asymmetries for the four neutrino types
are   in the relation:
\begin {equation}
A_{\nubar_e} < 
A_{\nu_\mu}  < 
A_\nu^{\rm 1D} < 
A_{\nubar_\mu}  <
A_{\nu_e} 
\end{equation}
where $A_\nu^{\rm 1D}$ is 
the asymmetry  predicted  in a one--dimensional approximation, that
is  approximately equal for all  neutrino types.

\item The  neutrino  type with the largest
asymmetry  is the $\nu_e$. The majority of these neutrinos
are  produced in the chain  decay $p \to \pi^+ \to \mu^+ \to \nu_e$,
and for  all these neutrinos the  contributions of the 
geomagnetic effects on the primary  particle and on the muon
add  to each other.

\item The east--west asymmetry for  $\nubar_\mu$ 
is smaller  that for $A_{\nu_e}$  but  larger than the 
1--D prediction.  This  is  a  consequence of the fact
that  approximately  one half of the $\nubar_\mu$'s  are
produced in  $\mu^+$   and   have a large asymmetry (approximately
equal  to $A_{\nu_e}$)
and for these neutrinos
the geomagnetic  effects on the primary  particles  and on the muons
add to each other.  The other half of the 
$\nubar_\mu$'s  is produced in  pion decay, and  will 
have  an asymmetry  close to what is  predicted in the 1--D  approximation.

\item Developing a  similar  argument 
we can conclude that the asymmetry for
$\nu_\mu$'s  will be  smaller that the 1--D prediction, because
for  approximately  half of these neutrinos
(those produced in the chain decay
$p \to \pi^- \to \mu^- \to \nu_\mu$)
 the effects of the bending of  the muons  gives
a  negative contribution to the asymmetry.

\item  The asymmetry for $\nubar_e$  will be the smallest one
since most of the particles  are produced in the decay of negative muons,
and  for  nearly all $\nubar_e$'s the effect of the bending of the muons
must be subtracted from the  asymmetry generated by geomagnetic effects  on 
the primary particles.
Numerically we  actually will find a 
  {\em  deficit} of east--going  $\nubar_e$'s, that is
an  asymmetry  that has  changed  sign. The effect of the muon  bending
overcompensates for the  effect of the primary flux.

\item  Even if  the detector  is  not capable to  distinguish 
neutrinos  from anti--neutrinos,  the  effect of the  magnetic
bending of the muons  is still  measurable.
For $e$--like events the net effect is an 
enhancement of  the  asymmetry with respect to the 1--D 
expectation:
\begin{equation}
A_e > A_e^{\rm 1D} 
\end{equation}
This  can be simply deduced as a consequence of three steps:
(i)  the effect of  muon magnetic  bending on the asymmetry
 is positive for
$\nu_e$'s and negative  for $\nubar_e$'s;
(ii)  in    the neutrino flux the $\nu_e/\nubar_e$  ratio
is  approximately unity (more accurately $\sim 1.2$,  reflecting the
$\pi^+/\pi^-$  ratio   in the final states of   proton interactions);
(iii)  the  cross section  for neutrinos  is larger than for 
anti--neutrinos.

\item  With  a  very similar  argument,
we can estimate that the asymmetry of $\mu$--like events must
be smaller  than the 1--D calculation:
\begin{equation}
A_\mu < A_\mu^{\rm 1D} 
\end{equation}
The inequality is opposite with  respect to
the  $e$--like case, since  the effect
of the  muon magnetic  bending on the asymmetry is negative  for
neutrinos  (with the  larger cross section)  and
positive for anti--neutrinos.
\end{enumerate}

\section {Numerical results  from a full 3--D calculation}

To go  beyond  this  qualitative discussion we have performed
a   full three-dimensional montecarlo calculation of the 
atmospheric  neutrino fluxes, including  the bending  in the 
geomagnetic field  of all charged  particles in the cosmic ray showers.
This   montecarlo  calculation   has   been   already described
in  \cite{geometry}, where it was used  to illustrate  
effects of a  three--dimensional  calculation
  on the   predicted  zenith angle distibutions
of $e$--like  and $\mu$--like events.
We refer to   \cite{geometry} for more  details on  the   calculation
method,
here we will  only briefly summarize the main points.
We have  used  a straightforward, direct method  (see the appendix for a 
comparison with  other approaches).
The  first step of the calculation is the generation of
isotropic fluxes of    cosmic  rays (we used the   results
of \cite{bartol} for the  energy  spectrum and   mass composition)
at ``the top of the atmosphere''  (at a radius  $R = R_\oplus + 80$~Km).
The geomagnetic  effects   on the primary flux are  computed
studying the past  trajectories of each  generated particle. 
 Primary particles on  forbidden 
trajectories  are rejected. 
Primary particles  on allowed  trajectories  are  propagated in
the  magnetic  field (the International Geomagnetic Reference
Field for the year 2000  \cite {igrf}). About 97\% of the particles  
interact in the air (the remaining  fraction  only grazes the Earth and
continues its  travel). The air
is described as  having a spherically  symmetric 
distribution $\rho(r)$    obtained  from a fit to the US standard atmosphere.
The final states   of the hadronic  interactions
was    generated using  the   model  (and  montecarlo algorithms) 
of Hillas \cite{Hillas}.
All secondary charged particles are propagated  along  curved  trajectories
in the  magnetic  field, and a  shower is generated   developing a
``tree'' with a standard  technique.
For  the propagation of muons  we have included  the  
(crucially important) energy loss  for ionization in the air
(and   the trajectory correctly takes into account the  variation of the 
muon momentum).
Energy  loss  was  neglected  for  all other particles.
Multiple  scattering  was neglected  for  all  particles.
When  neutrinos  are produced their trajectories
(simple  straight lines)  are  studied. 
Each  neutrino  either ``misses'' the Earth, or 
crosses  its  surface twice,
 first as   down--going 
and then as   up--going.   
At each intersection  its  position  and  direction of the neutrinos
is recorded. 
To compute the  azimuth of the  neutrinos, for  each  point on the surface of
the Earth we have defined  a  ``local'' system of  coordinates
as discussed in section  3: the $z$ axis  points up,
the    $x$ axis  toward 
magnetic  ``north'', and the $y$ axis  toward magnetic  ``west''.
In this system the  magnetic field  has  
 components  $(|B_{\rm hor}|, 0, B_{\rm vert}$).
The azimuth $\varphi$ (following the SK convention)  
is  defined as the direction where the particle is going:
a  particle with $\varphi = 90^\circ$ 
($270^\circ$) is traveling from east toward  west (from  west toward east).

The neutrinos  generated   with these algorithms  are distributed
over  the entire  surface of the Earth, with  important 
non--uniformities due to the  effects of the magnetic  field.
We have collected the neutrinos  in  five different regions
of equal area selected  according to the  magnetic latitude: 
an  equatorial region, two intermediate and two polar regions.
The results in the two  (north and south) polar  regions and
the two intermediate regions  are essentially undistinguishable,
and  here we present  only the results for the northern  regions.

Our results on the  azimuth distributions
are  collected in 6 figures (from fig.~3 to
fig.~8). In the  figures we 
plot  the  azimuth distribution of the event  rates,  obtained
after the convolution of the calculated fluxes with the model of the neutrino
cross section from \cite{LLS}.  We have
selected    neutrinos in the momentum interval:
 $p_\nu ({\rm GeV}) = [0.5, 3]$) and the zenith angle region
$\cos \theta_\nu = [-0.5, 0.5]$.
No  detection efficiency or experimental smearing
has been included.
For  each  of the  three regions of the Earth that we have considered,
we include two figures. The first figure contains   (in four separate  panels)
the  azimuth  distributions  of  the four neutrino  types.  The second 
figure  (in two  separate panels)   contains
the azimuth  distribution of  $e$--like and $\mu$--like events
(these distributions  are  obtained  simply summing together the results
for $\nu_e$ and $\nubar_e$  (or $\nu_\mu$ and $\nubar_\mu$).
The scale of the $y$  axis is absolute for all figures.

We have also performed the montecarlo calculation three times.

\begin{itemize}
\item [(i)] A first  calculation  (represented by thick histograms)
was  performed  using  the  ``full 3--D'' algorithms  descrivbed above, 
including the bending  of all secondary  charged  particles  in the showers.

\item [(ii)] A second  calculation (represented by thin histograms)
was   performed  to   reproduce  the  1--D  algorithms.  
The geomagnetic  effects  are calculated  for  the primary cosmic
ray particles  traveling outside the Earth's atmosphere
(exactly as in the previous case), but  all particles   travel along
straight line trajectories 
 for $r < R_\oplus + 80$~Km.). 
Also  all    final  state particles  are collinear  with the projectile
(or  parent)  particle.  This  is achieved 
modeling    the interactions  and     particle  decays 
exactly as  in the previous case,
(including therefore transverse momentum), 
and  performing as a last step a rotation  of all 
the 3--momenta  of the final  state particles 
so that  they  become   parallel  to the projectile
(for  interactions)  or  parent (for decays) particle.

\item [(iii)] A third  calculation  (represented  by  dashed  histograms)
was performed  neglecting
the  geomagnetic effects on the primary flux (therefore  considering
exactly isotropic  primary fluxes)  and  using the 1--D
algorithms  outlined in the previous point.
With these approximations the  azimuth angle distributions
{\em must} be  flat and independent from the  detector
position. The only non trivial  result
of the calculation is the absolute normalization 
of the  different rates.
Note that this normalization must also be  independent from the
geographycal  region considered.
\end {itemize}

Inspecting  the results of fig.~3  to fig.~8, we can make 
the following remarks.
\begin{enumerate}
\item  Calculation (iii)  (no geomagnetic  effects)
results as  expected in flat, detector position independent
azimuth  distributions.

\item Calculation (ii)  (1D with geomagnetic  effects included for the primary)
exhibits  as expected an east--west asymmetry, that is of approximately
the same  size for all four neutrino types.

\item  The taking into account of the effects of the 
bending in the magnetic  field  of secondary particles
(and also  of the primary  in the atmosphere)  in  calculation
(iii) is a non  negligible  correction. 

\item Comparing the  results of calculation (ii) and calculation (iii)
one can see that the effects  of the bending of secondary particles
result in an enhancement  of the  east--west asymmetry  for $\nu_e$,
and $\nubar_\mu$, and a suppression of the asymmetry for 
$\nubar_e$ and $\nu_\mu$.

\item The asymmetry for the combination  $(\nu_e + \nubar_e)$ is
also  enhanced, while the  
 asymmetry for the combination  $(\nu_\mu + \nubar_\mu)$ is
suppressed.
\end{enumerate}

In  summary we  can observe that the  expectations   of the qualitative
discussion of the previous  section have been confirmed  in a
preliminary, but detailed,  quantitative calculation.
The results  are  summarized in table~1, that gives the results
of the  asymmetries   obtained with  the calculation (ii)  (or 1--D)
and the calculation (iii)  (or 3--D), in the different
geographyical regions.
For the definition of the magnetic  latitude we have used
as  dipole  axis the one  that  corresponds to the leading  term in the
multipole expansion of  the IGRF field for the year 2000.
With  this   definition the SK  detector  has  a magnetic  latitude
of $27.08^\circ$ ($\sin \lambda_M^{\rm SK} = 0.455$) and is  
inside  the ``intermediate'' region in the northern hemisphere.

\begin{table}[hbt]
\caption{East--west asymmetry  for different event types, 
after  averaging  for the detector  position in  different 
 regions of the Earth. 
The   events are selected with the cuts  $E_\nu({\rm GeV}) = [0.5,3]$
and $\cos \theta_\nu = [-0.5,0.5]$. 
The two results are for a one--dimensional and
  a three-dimensional  calculation
that includes the bending of secondary  particles in the geomagnetic field. }
\begin {center}
\begin{tabular} {| l || c | c || c | c || c | c | }
\hline
Region & 
\multicolumn {2} {c ||} {Equatorial} & 
\multicolumn {2} {c ||} {Intermediate} & 
\multicolumn {2} {c  |} {Polar}  \\ 
 & 
\multicolumn {2} {l ||} {$\sin \lambda_M = [-0.2,0.2]$ } & 
\multicolumn {2} {l ||} {$\sin \lambda_M = [0.2,0.6]$ } & 
\multicolumn {2} {l  |} {$\sin \lambda_M = [0.6,1]$ }  \\
\hline 
 Event type &  $A$ (1D)   & $A$ (3D) &   $A$ (1D) & $A$ (3D) & $A$ (1D) &   $A$ (3D)    \\
\hline
 $\nu_e$        &    0.194 &    0.431 &    0.126 &    0.335 &    0.026 &    0.214 \\
 $\nubar_e$     &    0.189 &   $-0.057$ &    0.120 &   $-0.065$ & 
   0.012 &  $ -0.061$ \\
 $\nu_\mu$      &   0.180 &    0.065 &    0.115 &    0.028 &    0.011 &    0.002 \\
 $\nubar_\mu$   &   0.173 &    0.310 &    0.107 &    0.240 &    0.014 &    0.153 \\
\hline
 $e$           &    0.192 &    0.301 &    0.124 &    0.224 &    0.022 &    0.136 \\
 $\mu$         &    0.178 &    0.139 &    0.113 &    0.091 &    0.012 &    0.046 \\
\hline
\end{tabular}

\end{center}
\end{table}

Inspecting table~1, we can notice that as  expected  the  east--west asymmetry
depends  on the magnetic  latitude of the detector, and is strongest  near the
magnetic  equator,  where  all geomagnetic  effects are  enhanced.
In the 1--D calculation  the asymmetries  for the four  neutrino types  are approximately
equal, with small differences
  that  reflect  the  fact that the neutrino  spectra  produced
by a  primary of fixed energy  are not identical for the four  flavors.
The  full  3--D calculation  clearly 
exhibits the enhancements and  suppressions
predicted  qualitatively in section~\ref{sec:mu}.

\begin{figure} [t]
\centerline{\psfig{figure=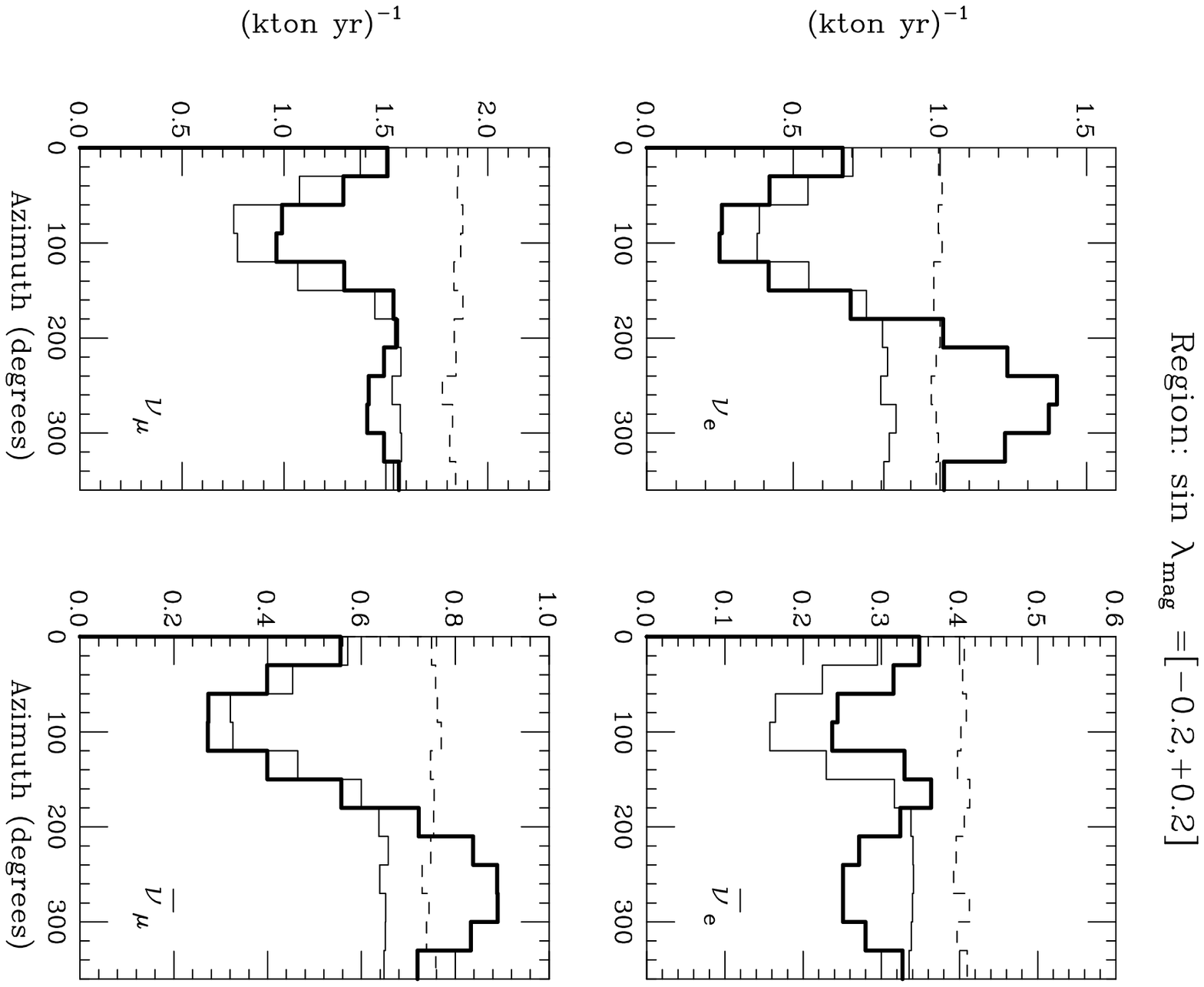,angle=90,height=18.0cm}}
\caption {Azimuth angle  distributions
of atmospheric  neutrino events.
The  distributions are  averaged 
for  detectors  uniformly distributed on  all positions  in the Earth 
magnetic  equatorial region ($ \sin \lambda_M = [-0.2,0.2]$).
The four panel  are for  neutrinos  of different  type.
The  three histograms  are  for: fully 3--D calculation
(thick), 1--D calculation (thin),  1--D without geomagnetic  effects
(dashed).  
\label{fig:azn3}  }
\end{figure}

\begin{figure} [t]
\centerline{\psfig{figure=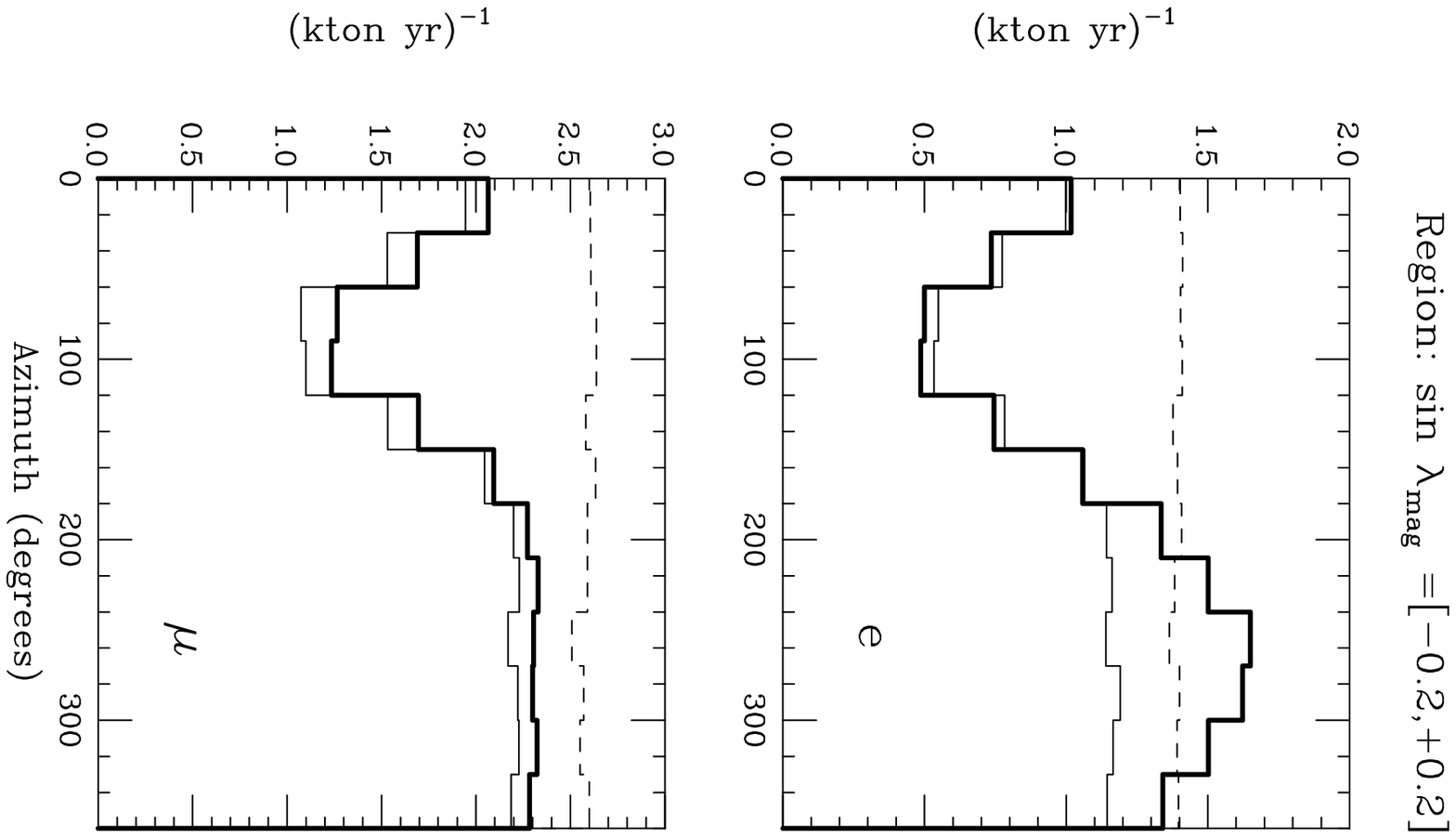,angle=90,height=18.0cm}}
\caption {Azimuth angle  distributions
of atmospheric  neutrino events.
The  distributions are  averaged 
for  detectors  uniformly distributed on  all positions  in the Earth 
magnetic  equatorial region ($ \sin \lambda_M = [-0.2,0.2]$).
The upper (lower)  panel  is for  $e$--like ($\mu$--like) events.
The  three histograms  are  for: fully 3--D calculation
(thick), 1--D calculation (thin),  1--D without geomagnetic  effects
(dashed).  
\label{fig:azl3}  }
\end{figure}

\begin{figure} [t]
\centerline{\psfig{figure=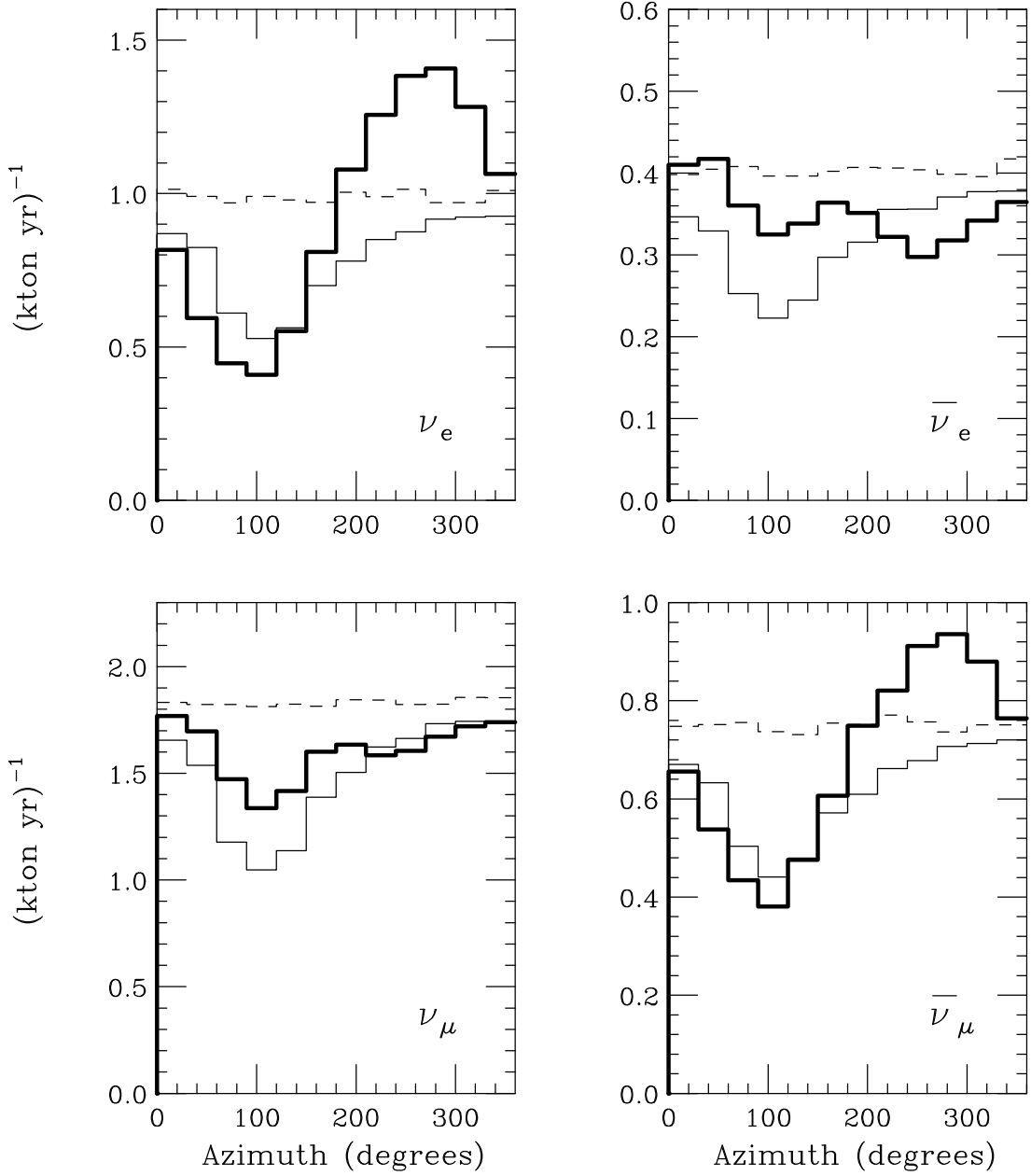,angle=90,height=18.0cm}}
\caption {Azimuth angle  distributions
of atmospheric  neutrino events.
The  distributions are  averaged 
for  detectors  uniformly distributed on  all positions on the Earth with 
magnetic  latitude  $ \sin \lambda_M = [0.2,0.6]$.
(The SK detector has  magnetic  latitude  $\sin\lambda_M^{\rm SK} = 0.455$)
The four panel  are for  neutrinos  of different  type.
The  three histograms  are  for: fully 3--D calculation
(thick), 1--D calculation (thin),  1--D without geomagnetic  effects
(dashed).  
\label{fig:azn2}  }
\end{figure}

\begin{figure} [t]
\centerline{\psfig{figure=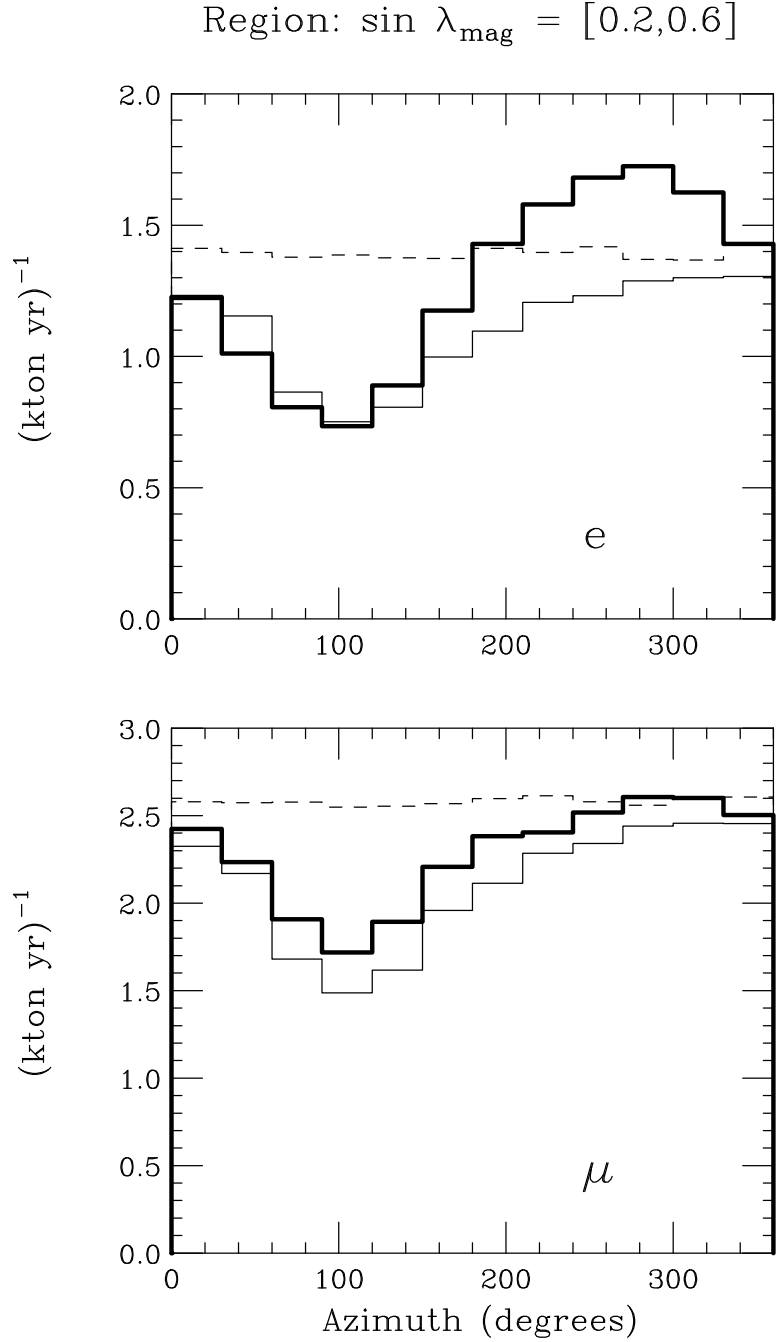,angle=90,height=18.0cm}}
\caption {Azimuth angle  distributions
of atmospheric  neutrino events.
The  distributions are  averaged 
for  detectors  uniformly distributed on  all positions on the Earth with 
magnetic  latitude  $ \sin \lambda_M = [0.2,0.6]$.
(The SK detector has  magnetic  latitude  $\sin\lambda_M^{\rm SK} = 0.455$)
The upper (lower)  panel  is for  $e$--like ($\mu$--like) events.
The  three histograms  are  for: fully 3--D calculation
(thick), 1--D calculation (thin),  1--D without geomagnetic  effects
(dashed).  
\label{fig:azl2}  }
\end{figure}

\begin{figure} [t]
\centerline{\psfig{figure=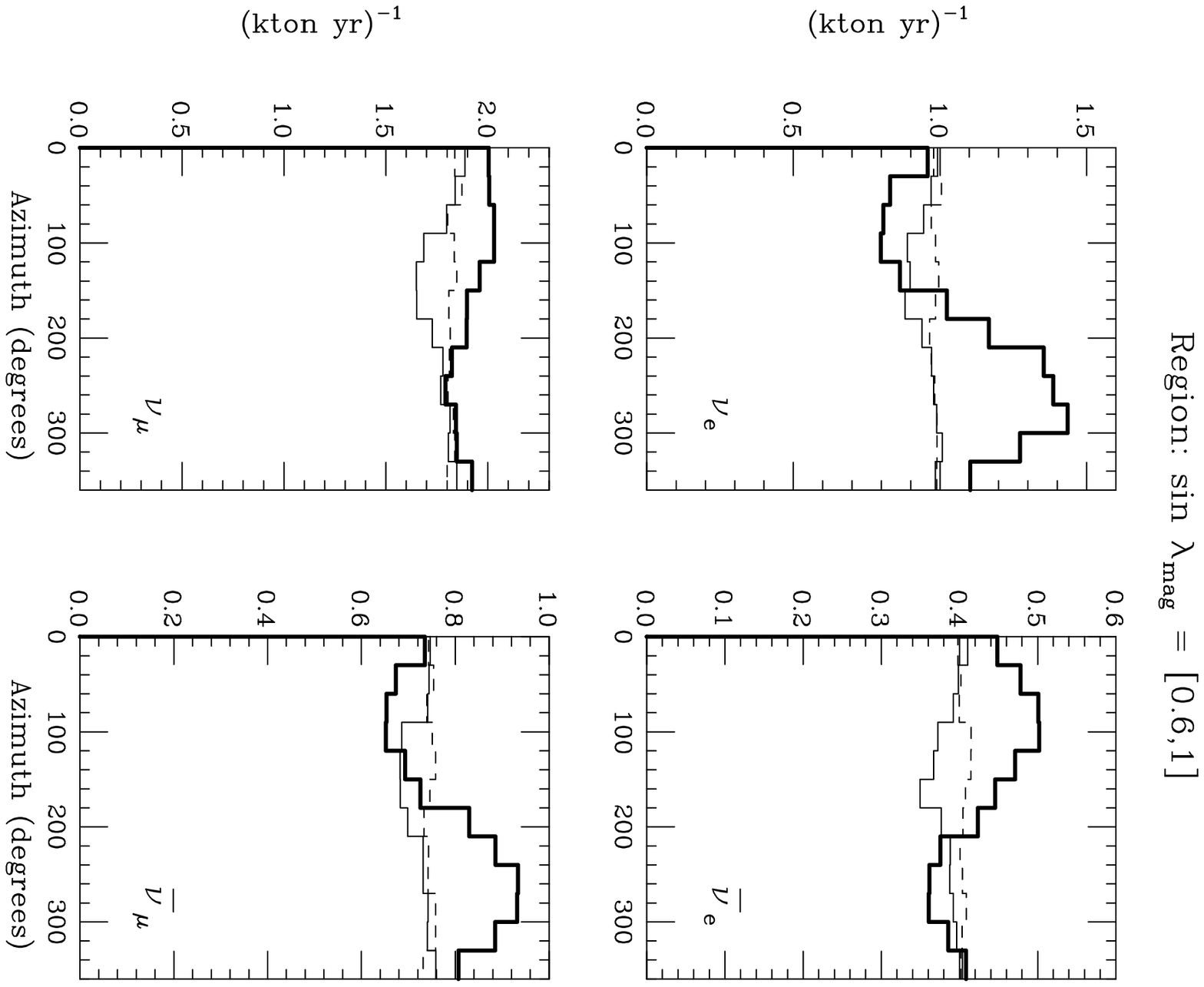,angle=90,height=18.0cm}}
\caption {Azimuth angle  distributions
of atmospheric  neutrino events.
The  distributions are  averaged 
for  detectors  uniformly distributed on  all positions  in the Earth 
magnetic north   polar  region ($ \sin \lambda_M = [0.6,1]$).
The four panel  are for  neutrinos  of different  type.
The  three histograms  are  for: fully 3--D calculation
(thick), 1--D calculation (thin),  1--D without geomagnetic  effects
(dashed).  
\label{fig:azn1}  }
\end{figure}

\begin{figure} [t]
\centerline{\psfig{figure=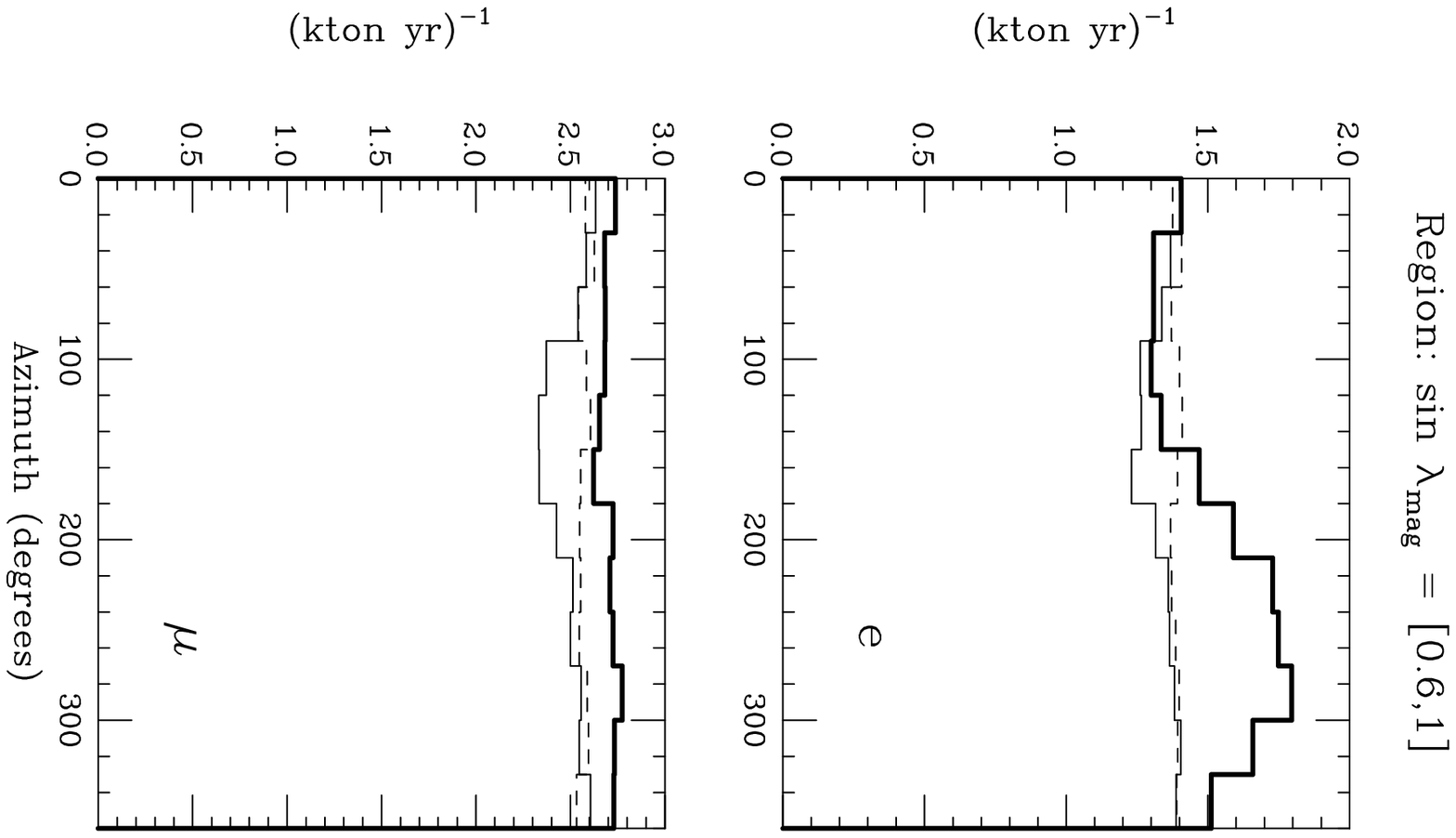,angle=90,height=18.0cm}}
\caption {Azimuth angle  distributions
of atmospheric  neutrino events.
The  distributions are  averaged 
for  detectors  uniformly distributed on  all positions  in the Earth 
magnetic north   polar  region ($ \sin \lambda_M = [0.6,1]$).
The upper (lower)  panel  is for  $e$--like ($\mu$--like) events.
The  three histograms  are  for: fully 3--D calculation
(thick), 1--D calculation (thin),  1--D without geomagnetic  effects
(dashed).  
\label{fig:azl1}  }
\end{figure}

\section {Summary and conclusions}

In this  work
we have  shown that  the azimuth angle distributions  of 
atmospheric  neutrinos  are  shaped  by the effects of the  geomagnetic
field on both the  trajectories of the primary cosmic rays, 
 and their  showers
in the Earth's atmosphere. 

In the existing (one--dimensional)  calculations   of the atmospheric
neutrino fluxes 
the effects of the magnetic  field on the shower development
are  ignored.  In  a  1--D   framework the asymmetries
of the  $\nu$  fluxes reflect
only the  highest suppression of the primary 
cosmic ray flux  when the   primary particles
travel towards west, and the predicted east--west asymmetry is
approximately equal for all  four neutrino types:
 ($\nu_e$, $\nubar_e$, $\nu_\mu$ and $\nubar_\mu$).

The situation changes when one  includes the 
bending  in the geomagnetic
field of  secondary particles (and in particular of muons) in the   cosmic  rays showers.
Positively and  negatively charged  particles  
are  curved  in  different  directions
and the resulting  effect on the east--west asymmetry 
 is  different for different
$\nu$ types. It is  an  enhancement   
for the neutrinos  that  are the product of
$\mu^+$ decay  ($\nu_e$, and $\nubar_\mu$)
and a suppression 
 for  the  products of
$\mu^-$ decay  ($\nubar_e$, and $\nu_\mu$).

For   detectors  like Super--Kamiokande that  cannot   separate
$\nu$'s and $\nubar$'s the detectable effect is  an enhanced
east--west asymmetry for the $e$--like events, and a suppressed asymmetry
for the $\mu$--like events,  as it is already observed \cite{SK-eastwest}
by Super--Kamiokande.
It is remarkable
that  the detctor   is able  to measure successfully 
 subtle  effects as those 
described here (and  without the need of a prediction).
In our  view this is  a  confirmation  of the 
 high  quality of the experimental data.

The  (partial)  failure of the existing calculations   in   the  prediction
of the east--west asymmetry is also  a  warning and an  indication that
the development of  more    accurate and  detailed  calculations of the
atmospheric  neutrino  fluxes is  needed. Several  groups \cite{future}
are  working in this  direction.
We note that a correct prediction of the east--west effects is possible 
only with a fully three--dimensional calculation.
In our  view the success of the  calculation presented  here 
in   reproducing the SK  results  is also 
a confirmation  of the
non--trivial  geometrical effects  connected
with 3--D effects  and the geometry of the neutrino
source volume discussed in \cite{fluka-3d} and in more
detail in \cite{geometry}.

The  understanding that the  present calculations  of atmospheric  neutrinos
do not  describe  correctly  the east--west asymmetry  can  have some
interesting  consequences  for  the   determination of the oscillation
parameters from the atmospheric  neutrino data
 (in  SK and also other experiments,
in particular Soudan). 
We note that  the  effects  we have been discussing  are 
obviously  especially important  for   horizontal neutrinos, and 
that the effects   have  different  sign  for   $e$--like and
$\mu$--like events.  
The shape  of the  zenith angle  distribution of the atmospheric neutrino
fluxes, in particular close to the horizontal plane,
 and the relation between the  muon and electron 
event  rates   are (at least in the opinion of this 
author)  the most  important 
problems  in the  calculation   of the atmospheric  neutrino fluxes.
As  an illustration, 
in the simplest (and  favored)     picture of  
$\nu_\mu \leftrightarrow \nu_\tau$  oscillations
for   values of  $\Delta  m^2$   in the range indicated  by the 
data 
and    the  range of $E_\nu$ characteristic of the contained  events,
the value of $\sin^2 2 \theta$ can be  reliably calculated
comparing  the ``up--going''  and   ``down--going rates'', with 
a relatively small systematic error  and    even for a  poor 
(or even incorrect)
estimate  of $\Delta m^2$. This  is possible  since
the ratio of the rates is   to a good  approximation
determined  only   the ``asymptotic  forms''  of  the 
oscillation  probability 
($P_{\nu_\mu \to \nu_\tau} = 0$  or $\sin^2 2\theta/2$) where the 
value of $\Delta m^2$  is  absent.
On the other hand  the  determination  of
$\Delta m^2$  requires  a  detailed  fit of  the zenith angle  distributions,
with the  region close to the horizontal playing  a crucial role.
The  (nearly exact) up--down symmetry   of the $\nu$--fluxes is of 
 little  help in this
case.  The  comparison  of the $e$--like and $\mu$--like rates   remains  as a
powerful  tool  to estimate  the suppression (or  enhancement) of 
 a flavor  type,
but of course as  discussed in this work,  for 
precise quantitative evaluations
of the parameters one  need to study with  great care the  
relations  between the fluxes of  different neutrino  types.

We postpone  to a future work  a  quantitative  estimate of
the    effects   of a full three--dimensional  calculation 
in the determination of the 
allowed region  in the neutrino oscillation  parameter space.

Other important sources of  uncertainties \cite{venezia,gaisser-taup99}
in the determination of the allowed region  for the oscillation 
parameters  are the  input primary spectra, where  very valuable new data 
became recently available \cite{Caprice,AMS,Bess},
and the modeling of hadronic  interactions.
New  data on this  problem  would also  be  of great value \cite{Harp}. 
A detailed  description of the neutrino  cross section is also
needed  to compute the event rates,  and
new data  would  be  very valuable  and benefit also
the long--baseline  programs.


\vspace{1.0 cm}

\noindent {\bf  Acknowledgments}
Special thanks to Takaaki Kajita for  very useful
  discussions  and encouragement.
I'm  also  grateful to Ed Kearns  for  discussions about the SK  data, 
and   gladly acknowledge  discussions with 
Giuseppe Battistoni,
Alfredo Ferrari,
Tom Gaisser,
and Yoichiro Suzuki.
This  work was  also  possible thanks to the 
precious  help of  Massimo Carboni, Kenji Kaneyuki and Atsushi Okada.

\newpage

\section*  {Appendix: 3--D calculations of atmospheric neutrinos}
 The first calculations
of the atmospheric  neutrino fluxes have been performed  in a one--dimensional
approximation, that is  assuming that the neutrinos are
emitted    collinearly with the primary particle.
This  allows  an enormous  reduction of the size of  a montecarlo
calculation, because only the (formally vanishingly small)  
fraction  of the  cosmic
rays showers  that   has  trajectories that intersect the detector
has  to be   simulated.

The difficulty of  a  a  full three--dimensional  calculation
of the neutrino fluxes has  been  discussed in \cite{fluka-3d} and
\cite{geometry}. 
In a  3--D  calculation  any shower  can produce a  neutrino
that   intersect  the detector  we are  considering, and therefore
all  possible  showers  have to be  studied. 
Since the  atmospheric  neutrinos 
are  generated 
quasi--uniformally over  the entire surface of the Earth
with a  total area  $5.1 \times 10^8$~Km$^2$, 
only a very small  fraction of these neutrinos
will  intersect a  dector  with an area  of order
$10^{-3}$~Km$^2$. 
Of course it  is possible  to consider  a   detector area
greatly enlarged  to  improve the statistical precision of the
calculation, however this  cannot  be done  arbitrarily because
in fact the  $\nu$ flux is not  exactly uniform,
but it does depend  on the detector  location.
This in a nutshell  is the   computational problem of
a 3--D  calculation.

There are  several  methods  that have been used  or
are currently  under study to overcome this  difficulty.

\subsection {Spherically  symmetric problem}
A useful  first step    is  to  consider  a 
 a simpler,  sperically symmetric  problem,  that  is obtained
neglecting all effects  of the  geomagnetic  field  both 
for the determination of allowed  and  forbiddden  trajectories,
and in the development of the showers. 
This  problem is  explicitely   spherically  symmetric:
all points on the surface of the Earth are  equivalent, and the
entire surface can be used  as the ``detector''.
This  calculation requires therefore the same  computer  power
of a 1--D calculation,   because 
essentially all produced  neutrinos
(all those that intersect the Earth's  surface)
can be collected and analysed 
to ``measure'' 
the $\nu$--fluxes in a montecarlo calculation.

A calculation along these  lines  has  been performed
by G.~Battistoni et al \cite{fluka-3d}.
This  calculation for the first time   demonstrated   the
existence of non--trivial  geomatrical  effects  
(see \cite{geometry} for more  discussion) due  to the 
spherical  geometry of the neutrino  source  volume.

\subsection   {Shower  translation algorithm}
Unfortunately a  calculation  performed  using 
the  approximation of a spherically symmetric Earth  is not adequate for
the  analysis  of experimental data.
The  largest effect  that is  missing is the effect of the
geomagnetic  effect  that  ``forbids''  low rigidity  cosmic rays from
reaching the vicity of the Earth.
If one  neglects  the  effects  of the geomagnetic  field on the 
shower development  it is possible to  ``translate'' a shower  from
a position  on the Earth  to   an arbitrary one, keeping the zenith angle
of the primary particle (and therefore the zenith angle of all particles)
as  constant  (this  of  course corresponds to  a   rotation
of  a shower seen as  ``rigid body'' around the  center  of the Earth, plus
an additional rotation   around the new  vertical axis).
A  single  shower  generated  by montecarlo can then  be  ``used''
many times  with enormous  saving of computer time. 
The effects of the geomagnetic  effects in  the determination
of the allowed  and forbidden trajectories can be  easily included, checking
that  if  a given  shower  position corresponds to an allower of  forbidden
primary  trajectory and rejecting  forbidden trajectories.
A  preliminary  calculation using this  method
for the three sites of  Kamioka, Soudan  and Gran Sasso  has  been
recently made available on  the web by Battistoni et al.
\cite{fluka-3d-web}.

\subsection{Weighted  technique}
If the effects  of the magnetic fields are also included 
in the development of the shower,  the problem  loses  all  symmetry
and one  has  to confront   the  computational  problem of
a calculation that is intrinsically very inefficient.
A solution of this  problem that is under   study
\cite{fluka-3d-future}, 
is  to use  a  weighted  technique, generating 
the cosmic  ray showers  with a strong  bias 
in position and  direction, so that  the neutrinos  produced are more
likely to arrive in the  vicinity of the  detector  in  consideration,
and   using a weight system to  estimate correctly the flux.
In this method     approximately  half of the  cosmic  rays interact
in a small region above the detector  to  simulate the down--going flux,
while  approximately one half are generated  over the rest of the Earth's
surface.  For  these  events the  direction of the shower  is  chosen
preferentially in a   cone  with an    axis  that  ``points''
toward the detector.
In this way it is possible to collect a   sufficiently large 
statistics of  neutrinos  in a relatively small area around the detector
site.

\subsection {Direct approach}
A  possible  solution  is   also to  neglect  any attempt 
at simplification, or  optimization, 
and  simply  generate   all cosmic showers
on the Earth   with  a  realistic  distributions in position
and   direction,  that  only takes into
account the geomagnetic  effects 
studying  the development of the
showers  in a  realistic  magnetic  field.
Such a  calculation   will produce a population of   neutrinos
that is  distributed   quasi--uniformally over the entire 
surface of the Earth, with  the   non--uniformities
representing the expected variations in intensity due to geomagnetic
effects,  that produce a larger (smaller)  flux
in the magnetic polar (equatorial) region.
In this  direct  (or brute force) approach 
 the  calculation is not performed for
a given detector position,   because all positions
on the Earth are treated ``democratically''.
This  is the technique  used in \cite{geometry} and 
also in this work.
Its  only  drawback is that   the results 
that can  be obtained in a reasonable time
integrating then neutrino  results over a large
area of the Earth.

\end{document}